\documentstyle[12pt]{article}
\newcommand{\pr}{\paragraph{}}
\newcommand{\be}{\begin{equation}}
\newcommand{\ee}{\end{equation}}
\newcommand{\bea}{\begin{eqnarray}}
\newcommand{\eea}{\end{eqnarray}}
\newcommand{\nd}[1]{/\hspace{-0.6em} #1}
\newcommand{\nk}{\noindent}
\begin{document}
\begin{titlepage}
\vspace{.1in}
\begin{flushright}
CERN-TH.6536/92\\
ACT-13/92 \\
CTP-TAMU-48/92 \\
\end{flushright}

\begin{centering}
\vspace{.1in}
{\large {\bf The String Universe: High $T_c$ Superconductor
or Quantum Hall Conductor? }} \\
\vspace{.2in}
{\bf John Ellis}, {\bf N.E. Mavromatos} and {\bf D.V.
Nanopoulos}$^{\dagger}$   \\
\vspace{.05in}
Theory Division, CERN, CH-1211, Geneva 23, Switzerland  \\

\vspace{.05in}
\vspace{.05in}
\vspace{.1in}
{\bf Abstract} \\
\vspace{.05in}
\end{centering}

{\small
Our answer is
the latter. Space-time singularities, including the initial
one, are described by world-sheet
topological Abelian
gauge theories
with a Chern-Simons term. Their
effective $N=2$ supersymmetry
provides an initial fixed point where the Bogomolny bound
is saturated on the world-sheet, corresponding to an extreme
Reissner-Nordstrom solution in space-time. Away from the
singularity
the
gauge theory has
world-sheet
matter fields, bosons and fermions,
associated
with the generation of target space-time. Because
the fermions are complex (cf the Quantum  Hall Effect)
rather than real (cf high-$T_c$ superconductors) the
energetically-preferred vacuum is not parity or
time-reversal invariant, and
the associated renormalization group flow explains
the cosmological arrow of time,
as well as
the decay of real or virtual black holes, with a monotonic
increase in entropy.}

\vspace{.1in}
\par
\vspace{0.4in}
\vspace{0.1in}

\par
\vspace{0.4in}

\begin{flushleft}
CERN-TH.6536/92 \\
ACT-13/92 \\
CTP-TAMU-48/92 \\
August 1992 \\
\end{flushleft}

\vspace{0.4in}

\noindent $^{\dagger}$ {\it Permanent address} :
Center for Theoretical Physics, Dept. of Physics,
Texas A \& M University, College Station, TX 77843-4242, USA,
and  \\
Astroparticle Physics Group,
Houston Advanced Research Center (HARC),
The Woodlands, TX 77381, USA.\\

\end{titlepage}
\newpage
\section{Introduction and Summary}
\pr
Condensed matter physicists have recently been fascinated with
conduction effects in two planar systems: Hall conductors
\cite{hall}
and
high-$T_c$ superconductors \cite{highsc}.
Both of these are described by QED in 2+1
dimensions interacting with matter that may be fermionic, bosonic or in
general anyonic. At the other end of physics, it has been realized
that Robertson-Walker-Friedmann cosmology \cite{aben}
and black holes in string
theory \cite{witt}
are described by coset Wess-Zumino models with, in the
interesting cases of 2-dimensional or spherically-symmetric
4-dimensional systems, an Abelian U(1) or O(1,1) gauge theory on the
2-dimensional world-sheet. For reasons which
we explained recently \cite{emnd}
it is useful to make a homotopic
extension of the world-sheet Abelian gauge theory action to a third
dimension, an idea familiar from the Wess-Zumino term. Confronted
with this convergence between the field-theoretical descriptions of
string cosmology and condensed matter systems, it is natural to ask
in more detail whether the string Universe bears a closer similarity to
a Hall conductor or to a high-$T_c$ superconductor.
\pr
    The answer is a Hall conductor, as we have already suggested
previously \cite{emnd}. Early
theories of high-$T_c$ superconductivity
suggested violations of parity $P$ and time-reversal
invariance $T$, and a zero-magnetic-field Hall effect
\cite{hlad,aitch}.
However, physical high-$T_c$ superconductors exhibit neither feature:
in particular, they conserve both $P$ and $T$ \cite{expt}.
On the other hand, the stringy
description of the neighbourhood of a space-time singularity involves
in an essential way an Abelian Chern-Simons term that violates both $P$
and $T$, as in the Hall effect at non-zero magnetic field \cite{hall}.
This
similarity between space-time singularities and Hall conductors
extends to the underlying infinite-dimensional $W$-algebras
\cite{whoW}. In the
case of singular stringy space-times, a $W_{1+\infty}$ algebra
ensures the maintenance of quantum coherence \cite{emn1},
whilst a wedge
subalgebra has been exhibited in connection with the non-singular
Laughlin wave-functions of the Quantum Hall
Effect \cite{trug}. As we show later,
this can be elevated to the full $W_{1+\infty}$ algebra if Hall
conductors with non-trivial topology are considered, such as annular
or doped conductors. The convergence of these physical descriptions
is supported by the observation that the $c=1$ string model, which
represents the spatially or temporally asymptotic limit of a full
stringy singularity, can be regarded as an incompressible quantum
fluid \cite{indians}, as can Hall charge carriers \cite{hall}.
\pr
Specifically, stringy
singularities are described on the world-sheet by an effective
Chern-Simons-Higgs theory which exhibits $N=2$ supersymmetry. This
means that they satisfy the Bogomolny bound \cite{bog}
and correspond
(in target space-time)
to
extreme Reissner-Nordstrom black holes \cite{reissn}.
It also means that they
correspond to the zero-field Hall effect, which can be realized
on a honeycomb lattice \cite{hlad}, but is
not seen in the known high-$T_c$
superconductors, as we have already commented. The enhanced symmetry
at the black hole core contains a $W_{1+\infty} \otimes
 W_{1+\infty}$
algebra, whose breakdown away from the singularity to the
coherence-preserving $W_{1+\infty}$ algebra is accompanied by the
appearance of discrete massless modes \cite{emntop}.
Similar states should appear in
analogous Hall conductors.
\pr
    The $N=2$ zero-field Hall
    effect theory is a fixed point of the
renormalization group flow. The $T$- and $P$-violation in the
generic Hall conductor correspond to the irreversibility of black
hole decay and the arrow of cosmological time. We identify the Hall
fraction with the level parameter $k = 9/4$ of the stringy black hole
in the absence of additional matter. Hall conductors with different
fractions can be understood as conductivity plateaux in the Laughlin
hierarchy \cite{laugh}, and
the expanding Universe can be regarded as sliding down
this hierarchy \cite{aben}, with
the corresponding creation of matter degrees of
freedom.
\pr
    Corresponding to the above paragraphs, we discuss planar
electron systems in section 2, the correspondence of a Hall
conductor to a Minkowski black hole in section 3, and the r\^ ole
of $N=2$ supersymmetry and
the renormalization flow in
section 4. Section 5 develops
further the isomorphism between Hall conductors
and Minkowski black holes, and finally section 6
correlates the irreversibility of the
renormalization group flow, $P$ and $T$ violation in the
Hall vacuum, the cosmological arrow of time, black hole decay
and microscopic entropy growth.
\pr
\section{Planar Electron Systems}
\pr
    Let us review the general features of such systems, emphasizing
those characteristics relevant for our purposes. The action for
$QED_3$ with fermionic matter is
\be
   S= \int d^3x [-\frac{1}{4}F_{\mu\nu}F^{\mu\nu} +
   {\overline \psi} (i\nd{\partial} +
e \nd{A} ) \psi + \kappa \epsilon_{\mu\nu\rho} A_{\mu}\partial _{\nu}
A_{\rho} +...]
\label{one}
\ee

\nk where
we have allowed for the possibility of a Chern-Simons term
which violates $P$ and $T$ but not $PT$. Such a term is absent from
the fundamental Lagrangian, but can be generated by matter loops, for
example non-relativistic electrons via a Pauli magnetic moment term
\be
       S_{Pauli}=\frac{ge\hbar}{4Mc} \sigma B(x)   \qquad ;
\qquad \sigma = spin-matrix
\label{two}
\ee

\nk calculated in the adiabatic limit. Note that the effective action is
supersymmetric if the gyromagnetic ratio $g$
is $2$ \cite{gyro},
as expected for
elementary Dirac charge carriers.
\pr
    The fermionic current obtained from (\ref{one}) using the classical
equations of motion is transverse to the direction of the applied
electromagnetic field,
\be
J_{\mu}= 2\kappa \epsilon _{\mu\nu\rho}\partial _\nu A_\rho
\label{tria}
\ee

\nk so the charge density is proportional to the applied magnetic field,
and the transverse Hall conductivity is
\be
       \sigma _{xy} = \frac{\partial \rho }{\partial B}
\label{tessera}
\ee

\nk The $P$- and $T$-violation
inherent in the action (\ref{one}) can show up as a
non-zero current in the presence of a non-zero magnetic field - the
usual Hall effect which occurs within separated Landau levels. However,
there can also be a zero-field Hall effect in the form of a non-zero
transverse conductivity (\ref{tessera}) even in the limit $B \rightarrow
 0$, when the
Landau levels become degenerate. This is in principle a possibility
in planar superconductors, thanks to the Meissner effect.
\pr
    High-$T_c$ superconductors are planar, and it is natural to enquire
whether they exhibit such $P$- and $T$-violation. Many theories of
high-$T_c$ superconductivity exploit the existence of excitations with
fractional statistics, anyons, in planar systems \cite{anyons}.
When starting from
fermions or bosons, the description of anyons requires the
introduction of a statistical Abelian gauge field $a$, which will in
general have its own Chern-Simons term, as well as a mixed
$a-A$ Chern-Simons term in the effective action. Such anyonic
theories of high-$T_c$ superconductivity do indeed violate $P$ and $T$
in general, and predict a zero-field Hall effect. Neither of these
features are seen in known high-$T_c$
superconductors \cite{expt}, and such anyonic
theories are therefore ruled out. $P$ and $T$ are, however, conserved
if the charge-carriers are real fermions, which is predicted in
one microscopic theory of high-$T_c$
superconductivity \cite{dm}. This theory
also predicts successfully the ratio of
the gap $\Delta$ to $T_c$ \cite{dm,adm}.
\pr
    Stringy singularities clearly resemble more closely the Hall
conductor mentioned in the next-to-last paragraph: they are described
(see the next Section) by Abelian world-sheet gauge theories with a
(dimensionally-reduced) Chern-Simons term interacting with
complex fermions. Anyons play no apparent role in the known singular
string solutions, and there is no analogue of the statistical gauge
field on the world-sheet. Moreover, just like the Hall electrons in
each Landau level, which form an incompressible quantum fluid, there
is an equivalent structure
in the $c=1$ string model \cite{indians}, which describes
the spatially- or temporally-asymptotic form of the singular stringy
solution.
\pr
    This physical similarity is reflected in the underlying
group-theoretical structure. It has been known for some time that the
$c=1$ model and stringy black holes possess a global $W_{1+\infty}$
symmetry on the world-sheet \cite{bakir,wu},
which is elevated to a local symmetry in
target space-time \cite{emn1,emnsel}.
This W-algebra contains an infinite Cartan
subalgebra, whose associated charges constitute W-hair for the black
hole which maintains quantum coherence \cite{emn1}.
A wedge subalgebra of a
similar W-algebra has recently been identified in the integer Quantum
Hall Effect \cite{trug}.
However, because of physical regularity conditions on
the Laughlin wave functions, the full W-algebra has not been
obtained. The requirement of
regularity would be relaxed in planar Hall
conductors with non-trivial topology, such as an annulus. Indeed, as
we shall see in more detail later on, an annular Hall conductor is an
accurate model of a black hole in string theory. The latter is
described by a spike-antispike configuration on the world-sheet, and
these defects correspond to the inside and outside of the annulus. A
more general multi-spike-antispike configuration, which describes a
generic foamy state of target space-time with many microscopic black
holes, can in principle be realized by doping a Hall conductor. The
corresponding generalization of the Laughlin wave function need not
be regular at any of the doping sites, and hence would
realize
a similar
mathematical structure as well as underlying quantum fluid picture.
\pr
\section{Correspondence to Minkowski Black Hole}
\pr
    We now demonstrate in more detail the connection of a Hall
conductor to a Minkowski Black hole in string theory.
We first recall
that a Minkowski black hole is described on the world-sheet by an
$SL(2,R)/O(1,1)$ Wess-Zumino coset model \cite{witt},
which in the neighbourhood of
the space-time singularity becomes a $O(1,1)$ topological gauge theory
coupled to matter fields \cite{witt,eguchi}
:
\be
S_{eff}=  -\frac{k}{4\pi}
   \int d^2z \sqrt{h} h^{ij}D_ia D_jb
  +i\frac{k}{2\pi} \int d^2z w\epsilon ^{ij}(F(A))_{ij} + ...
\label{threon}
\ee

\nk Here $w$ is the degree
of freedom describing the singularity, the
$D_i$, $i$ = 1,2 are gauge covariant derivatives, $(F(A))_{ij}$ is the
field strength of the Abelian $O(1,1)$ gauge potential, the bosonic
fields $a$, $b$ are $SL(2,R)$ coordinates, and the dots represent
higher-order terms in the expansion around the singularity, which is
represented by a spike on the world-sheet \cite{emnd}.
\pr
    The correspondence to the planar electron systems of the previous
section, which are described by
Abelian (electromagnetic) gauge models coupled to fermions, becomes
clearer if we regard the effective action (\ref{threon})
as the adiabatic
approximation to a three-dimensional gauge theory.
This limit is
particularly relevant in view of the relation given in
ref. \cite{emnd} of space-time foam to multi-defect configurations on
the world-sheet. The present-day physical phase described by a plasma
of Minkowski black holes arises at low ``temperature'', defined to be the
homotopic parameter of the third dimension compactified on a circle of
radius $\beta$ = $ T^{-1}$. The three-dimensional effective gauge
theory whose constant-``temperature''
sections define the conformal WZ
model corresponding to a Minkowski black hole is a scalar
Higgs-Chern-Simons gauge theory \cite{jackiw}.
Defining
\be
\Phi \equiv \left( \begin{array}{c}
\nonumber  a + b \\
           a-b        \end{array}\right)
\label{triadio}
\ee

\nk it can be written in the form
\be
\frac{{\tilde k}}{4\pi}
\int _{\Sigma ^2 \times S^1} d\tau d^2z \sqrt{h}
h^{ij} D_i {\overline \Phi } D_j \Phi
-i \frac{{\tilde k}}{2\pi} \int d \tau d^2z
\epsilon _{\mu\nu\rho} A(\Sigma ,\tau)_\mu \partial _\nu
A(\Sigma ,\tau)_\rho
\label{triatria}
\ee

\nk where ${\tilde k} \equiv \frac{k}{\beta}$, and
${\overline \Phi} \equiv \Phi \sigma _{3}$,
with $\sigma _3 $ a $2 \times 2 $ Pauli matrix.
One can easily obtain
equation (\ref{threon}) by making a dimensional reduction with respect to
$\tau$ and eliminating the $A_\tau$ component of the gauge potential
via its
equation of motion. The theory (\ref{triatria}) can then be written in
terms of a single complex field
by identifying
$ {\overline \Phi}$ with the
complex conjugate of $\Phi$.
\pr
To complete the
model (\ref{triatria}) we must specify the effective scalar
potential. Since the approach to the singularity is equivalent to a
symmetry-restoration process, as described in ref. \cite{emntop}, the
effective potential should vanish when $\Phi  = 0$. It should also
have a symmetry-breaking minimum at some non-zero $|\Phi|=  V$. On the
other hand, if we want the three-dimensional quantum theory to be
well-defined, we should require it to be renormalizable. The most
general potential obeying all these requirements is
\be
V(\Phi )=\frac{\alpha e^4 }{{\tilde k}^2 } (|\Phi |^2 -V^2 )^2
 |\Phi |^2
\label{triatessera}
\ee

\nk This
corresponds to the Bogomolny limit of interest to us if $\alpha
= 1/4\pi ^2 $ \cite{jackiw}.
As already mentioned, in the case of black holes,
the singularity corresponds to
the symmetric state $\Phi$ = 0, whilst the horizon corresponds to
$\Phi =1$ , and hence $V = 1$
in the effective potential (\ref{triatessera}).
\pr
We now observe that
the Minkowski black hole is a static solitonic
configuration of the model (\ref{triatria},\ref{triatessera})
which satisfies the classical
Euler-Lagrange equations:
\bea
\nonumber D_1 \Phi &=& \mp iD_2 \Phi \\
e\epsilon ^{ij}\partial _i A_j \equiv e B &=& \pm
\frac{(8\pi e^2)^2}{ {\tilde k}^2}|\Phi |^2(1-|\Phi |^2 )
\label{triapente}
\eea

\nk It was shown in ref. \cite{jackiw} that
the equations (\ref{triapente})
possess
topologically stable solutions for which $|\Phi| \rightarrow 1  $
at large
distances, and the corresponding magnetic flux is quantized. However,
these solutions are not consistent with the stereographic embedding
\be
 |z|^2 =-uv
\label{triaexi}
\ee

\nk
of the world-sheet into space-time, which leads to the target-space
Minkowski black hole metric
\be
ds^2_{target}=\frac{1}{1-uv}dudv
\label{triaepta}
\ee

\nk We see in (\ref{triaexi}), (\ref{triaepta}) that
the origin of the stereographically-projected
 world-sheet corresponds to the horizon $uv  = 0$ of the black hole,
where $|\Phi|= 1$, since the $SL(2,R)$ coordinates
obey $ab + uv = 1$.
The embedding (\ref{triaexi}) actually describes only the
interior of the horizon of the Minkowski black hole. Thus we can
without loss of generality place the anti-spike describing the
space-time singularity at the spatial world-sheet point at infinity.
The Minkowski black hole is therefore described by a solution of the
Higgs-Chern-Simons
equations (\ref{triapente}) in which the Higgs field $\Phi$
approaches the symmetric vacuum asymptotically at large distances.
In this case, the
equations (\ref{triapente}) possess non-topological soliton
solutions, for which the magnetic flux is not quantized, but
continuously varying. At the singularity, the flux vanishes as argued
in ref. \cite{emnd}
on the basis of a trivial $w$-integration.
\pr
    An important aspect of this Bogomolny limit is that the mass of the
elementary excitations of the system equals their electric charge. This
can be seen as follows: due to the Chern-Simons term, an object carrying
a magnetic flux $\Phi _M$ also carries a non-zero electric charge $Q$:
\be
      Q=-\frac{{\tilde k}}{2\pi} \Phi _M
\label{triaokto}
\ee

\nk The energy $E$ of a non-topological soliton is then given by
\be
    E = e|\Phi _M | =\frac{2\pi e}{{\tilde k}}|Q|
\label{triaennea}
\ee

\nk which
is the lower bound of the general energy relation\cite{jackiw}
\be
E \ge e |\Phi _M |
\label{triadeca}
\ee

\nk which follows from the general expression for the energy in the model
(\ref{triatria},\ref{triatessera}) without using
its equations of motion (\ref{triapente}). Hence the
Minkowski black hole is of extreme
Reissner-Nordstrom type \cite{reissn}.
\pr
As already mentioned,
in a previous paper \cite{emnd}
we showed that the magnetic flux $\Phi _M$ vanishes
at the singularity, as the result of a trivial integration over the
$w$ variable. However, that argument does not apply away from the
singularity, where the action is not of pure Chern-Simons type, but
has
extra terms that can be described collectively as the effective
potential discussed above. This is why
the flux can in general be non-zero
away from the singularity, leading to non-topological solitons with
continously-varying non-zero fluxes, interpretable as possible masses
of Minkowski black holes.
\pr
\section{Supersymmetry and Renormalization Group Flow}
\pr
  We have already recalled that the region around the singularity can
be described by a topological field theory (TFT). This can
be put
in a form with twisted N=2 supersymmetry, in which the supersymmetries
become fermionic BRST gauge symmetries $F$, and the fermions become
ghosts. As was pointed out in ref. \cite{eguchi}, fixed
points in the action
of $F$ dominate the path integral. An example is the region around a
singularity in two-dimensional space-time, which is described by a
twisted N=2 supersymmetric Wess-Zumino (SWZ) model. This leads to a
super $W$-symmetry at the singularity, which contains an enhanced
$W_{1+\infty} \otimes W_{1+\infty}$
bosonic symmetry at the core of the black
hole \cite{emntop}.
This enhanced symmetry is broken as one moves away from the
singularity, and the symmetry breaking is accompanied by the appearance
of discrete massless modes \cite{emntop}.
However, although the topological
nature of the theory is lost away from the singularity, its
supersymmetry is maintained by undoing the twist present at the
core. The fermionic fields that are ghosts at the singularity
become the usual fermionic partners of N=2 supersymmetry when the
model is untwisted, and the transformations $F$ have no fixed points
away form the singularity.
\pr
We now reconsider these ideas from a three-dimensional point of view.
Consider the region close to the core of a two-dimensional black hole,
where the physics is described by an SWZ model \cite{eguchi,noij}
\be
   S_{SWZ}=S_{WZ} + \frac{i}{2\pi}
   \int d^2z [ \Psi D_{{\bar z}} \Psi +{\overline \Psi } D_{z}
\Psi ]
\label{swz}
\ee

\nk and the fermions of the coset
model have been written in the matrix
form
$\Psi =\left( \begin{array}{c}0 \qquad  \psi \\
         \psi ^{*} \qquad 0         \end{array} \right)$ and
$ {\overline \Psi } =  \left( \begin{array}{c}0 \qquad {\overline
\psi} \\{\overline \psi}^{*} \qquad 0         \end{array} \right)$.
In the case of the Minkowski
black hole, the explicit N=1 supersymmetry can be enhanced to a higher
N=2 supersymmetry by imposing a  GSO projection \cite{noij}.
Since the
effective Higgs-Chern-Simons theory is known to have such an extended
N=2 supersymmetry \cite{lee}, we
assume that this is done before the homotopic
extension to the three-dimensional manifold.
\pr
Under this assumption, we rewrite the action (\ref{swz})
using a Dirac-like
notation for the fermions:
$\chi \equiv \left(
\begin{array}{c} \psi \\
{\overline \psi }\end{array} \right)$
so that :
\be
          S_{SWZ}^{fermion} =\frac{i}{2\pi} \int d^2z
{\overline \chi} \nd{D}(A) \chi
\label{dirac}
\ee

\nk As in
the purely bosonic case, the two-dimensional fermion-scalar theory
can be interpreted as the adiabatic approximation to a three-dimensional
 model, after using the equation of motion for the time component of
the gauge field. The homotopically-extended model is then an N=2
supersymmetric Abelian Chern-Simons-Higgs model with a
symmetry-breaking
vacuum. This we can interpret as a renormalization group
fixed point of a more general model of a charged scalar field coupled
to a Chern-Simons gauge field \cite{kaz}:
\be
\frac{1}{2}\epsilon^{\mu\nu\rho} A_\mu \partial _\nu A_\rho
+ |D(A)_\mu \phi |^2 + {\overline \chi} \nd{D}(A) \chi
+\alpha {\overline \chi} \chi \phi ^* \phi - h (\phi ^* \phi )^3 +...
\label{cp}
\ee

\nk
where the dots denote terms that are irrelevant in the renormalization
group sense, and play no role in our arguments. Also, we did not
write explicitly a quartic interaction because, although dominant in
the infrared limit, it does not move the fixed point \cite{kaz}.
\pr
The $\phi$ and $\chi$ fields can develop dynamical masses in the model
(\ref{cp}), via a
three-dimensional Berezinsky-Kosterlitz-Thouless (BKT)
phase
transition \cite{kt}
which can lead to superfluidity. However, we will argue that
this is not the case for the Minkowski black hole, which resembles a
Hall conductor rather than an anyon superfluid. To see this, we recall
the mass generation mechanism in the model (\ref{cp}): the nonlinear
interactions among the $\phi$ fields can be linearized in the
Hartree-Fock approximation by replacing pairs $\phi^*$$\phi$ by their
vacuum expectation value $ < \phi^*  \phi >$. This produces a theory with
a dynamical effective potential of the form (\ref{triatessera})
without a local order parameter. Dynamical
mass generation can be studied in this model with the aid of the
large-N expansion for the number of scalar fields \cite{dm},
although there is
just one scalar field in the physical case. The large-N expansion gives
a correct qualitative description of the gap, but one must go beyond
this approximation to see the BKT nature of the transition. The result
of the analysis depends on the values of the respective couplings.
Among the leading-order renormalization group fixed points there is
one with N=2 supersymmetry given by
$\alpha=3e^{ 2}, h=e^{ 4}$.
This fixed point is infrared stable \cite{kaz},
implying that supersymmetry
is an asymptotic symmetry of the system. This observation confirms
our physical intuition about the importance of the Bogomolny limit
for black hole physics. Viewed from three dimensions, the
underlying theory has a cut-off parameter whose variation drives it
to a non-trivial N=2 SWZ model describing matter interacting with
a Reissner-Nordstrom black hole. Its extremality is intimately
connected with supersymmetry \cite{oliv}. In view of this
supersymmetry, this fixed point separates phases in which either
both fermions and scalars have identical non-zero masses, or
both are massless. Clearly, the space-time singularity corresponds
to the massless phase, whilst the non-singular part of the
space-time, including the horizon, corresponds to the massive phase.
\pr
The three-dimensional system violates both time-reversal T and parity P
in the massive phase. This is because when one integrates out a
Dirac fermion with mass
\be
    m_f = \langle \phi ^* \phi \rangle 3e^2
\label{massf}
\ee

\nk one generates a Chern-Simons term with coefficient
$sgn(m_f)\frac{1}{8\pi}$, which
is independent of the magnitude of the mass \cite{redlich}. This
is unable to
cancel the bare Chern-Simons coefficient in general, and certainly
not in the adiabatic limit where the bare coefficient vanishes.
The induced $T-$ and $P-$ violation are physical motivations for our
identification of black holes with Hall conductors rather than
high-$T_c$ superconductors. The latter do not show any experimental
signature of T and P violation \cite{expt}, a
fact attributed by one of us (N.E.M.)
and N. Dorey \cite{dm}
to the observation that defects in high-$T_c$
superconductors correspond to an even number of species of
real fermions, for which the induced
Chern-Simons coefficient vanishes. In our case, the non-compact
nature of the coset and supersymmetry give us a complex fermion
representation that induces a non-zero coefficient.
The consequent T violation
implies irreversibility in the flow of the homotopic scale or
cut-off parameter, corresponding to the irreversibility of black
hole decay. Moreover, in view of the embedding (\ref{triaexi}) of the
world-sheet into space-time, the induced $P-$ violation corresponds
to an ``arrow'' in the target time direction. This observation
leads to the cosmological arrow of time, when applied to a
space-time with a cosmological singularity, as we discuss
in more detail in section 6.
\pr
The appearance of parity violation in the string vacuum does not
contradict any no-go theorems \cite{col}.
Spontaneous parity violation is
forbidden in vector-like local field theories \cite{vafwitt},
and in three-dimensional
gauge theories with fermions in real representations of the Lorentz
group SO(2,1) spontaneous or dynamical parity violation is
energetically disfavoured if there is an even number of fermion
flavours. However, these arguments do not apply when fermions
belong to complex representations of SO(2,1), and spontaneous
parity violation can indeed be energetically preferred in theories
with an odd number of complex fermions \cite{redlich},
as is the case in the
SWZ model of interest here.
\pr
It is instructive to make connections with the description of
target-space black holes as world-sheet spikes and vortices. At
the horizon, the composite scalar field in the effective gauge
theory description acquires a non-trivial vacuum expectation
value $< \phi^*  \phi > \ne 0$. Small fluctuations about this
non-trivial minimum in the effective potential can be fermionized
as follows. As already mentioned,
the horizon corresponds to the limit $uv \rightarrow
0$,
where the coefficient of the Chern-Simons term vanishes.
Due to supersymmetry, there is an induced fermion mass that
becomes infinite in this limit. Integrating out the gauge
field in this non-trivial vacuum, one easily obtains
\be
L_{eff} = -(\phi ^* \partial _\mu \phi + {\overline \chi}
\gamma _\mu  \chi )^2 + i {\overline \chi} \nd{\partial} \chi
+(scalar-sector)
\label{fr}
\ee

\nk Redefining the fermion fields by \cite{ruiz}
\be
\lambda \equiv \phi \chi \qquad : \qquad
{\overline \lambda }
\equiv \phi ^* {\overline \chi}
\label{redf}
\ee

\nk and
using the non-zero vacuum expectation value $< \phi^*  \phi > \ne 0$,
we find a massive 2+1-dimensional Thirring model
with an attractive four-fermion interaction:
\be
   i{\overline \lambda} \nd{\partial} \lambda
   -\frac{1}{4}({\overline \lambda}
\gamma _\mu \lambda )^2
\label{thir}
\ee

\nk Upon the dimensional reduction corresponding to the adiabatic
limit, this model becomes after bosonization \cite{colthir}
a 2-dimensional
sine-Gordon model which describes spikes on the world-sheet.
\pr
A similar mechanism operates at the singularity, which corresponds
to an antispike in the world-sheet picture. The bare Chern-Simons term
is non-vanishing at the singularity, and is the dominant term in a
derivative expansion, because of the short-distance nature of the
problem. It can therefore be fermionized by infinite-mass fermions,
which add an extra effective flavour to the Thirring model
(\ref{thir}),
that yield after bosonization \cite{colthir}
sine-Gordon terms that describe the
world-sheet antispike corresponding to the
singularity \cite{emnd}.
\pr
\section{Isomorphism with the Quantum Hall Effect}
\pr
We now develop the isomorphism of
the
above effective gauge model description
of a Minkowski black hole with the formalism for a Hall
conductor. We recall that the Hall effect has long been
described using a $\sigma$-model \cite{pruisk}
with a Chern-Simons term, in which the longitudinal
and transverse conductivities $\sigma _{xx}$, $\sigma _{xy}$,
are treated as couplings subject to renormalisation, the
conductivity plateaux corresponding to vanishing $\beta$-functions
and hence conformal invariance of the $\sigma$-model.
This approach has recently been extended from the integer to the
fractional quantum Hall effect \cite{lutk}, exploiting better
the complex $SL(2,Z)$ duality symmetry of the effective $\sigma$-model.
\pr
We have already touched on similarities of the effective gauge theory
description of the black hole WZ coset model to the physics
underlying the quantum Hall effect.
The
bosonic fields $(a,b)$
in the Lagrangian
(\ref{triatria})
have charges coupled
to the gauge field $A_\mu$, and the corresponding
current $\delta S_{eff} / \delta A_\mu $ can easily
be found by integrating out the scalar fields in the massive
phase which corresponds to space-time \footnote{We recall
that the massless phase corresponds to the
singularity, where the gauge field-theory is purely
topological.}. It is evident from the effective potential
(\ref{triatessera}) that the scalar fields acquire masses
of order $\mu = \frac{e^2}{{\tilde k}}$ near the symmetric
vacuum, where ${\tilde k}$ is the coefficient of the
Chern-Simons term. In the adiabatic limit, the homotopic
scale $\beta \rightarrow \infty $, hence ${\tilde k} \propto
\frac{1}{\beta} \rightarrow 0 $ and the scalar mass $\mu \rightarrow
0$. Thus the effective theory is well approximated by one-loop
graphs that yield Maxwell terms for the gauge field
\cite{polyakov}
\be
 S_{gauge}=\int d^3x[\frac{1}{24\pi |\mu |}
 (F_{\mu\nu}(A))^2  + \kappa A \wedge F(A) +
{\overline  \chi} (i\nd{\partial }
+ e\nd{A} )\chi + ...]
\label{integr}
\ee

\nk Notice the opposite sign of the induced Maxwell term
as opposed to the standard electrodynamics.
The fermion current then exhibits a transverse Hall form.
Summing over all ``times'' to obtain the two-dimensional
current
\be
{\hat J_i}=\int _{0} ^{\beta} j^{\chi}_i =\sigma ^{xy}_{ij}
 \partial _j w
\label{adcur}
\ee

\nk we find
\be
         \sigma ^{xy}_{ij} = -\epsilon _{ij}
         \frac{ke^2}{2\pi}
\label{hall}
\ee

\nk for the
transverse conductivity.
\pr
We see immediately from (\ref{hall}) that {\it the WZ coefficient
$k$ is the Hall fraction}. The Minkowski black hole case therefore
corresponds to a {\it fractional charge $\frac{9}{4}e$ } in the
language of the Hall effect.
\pr
As we have mentioned in section 2, a wedge
subalgebra of $W_{1+\infty}$ has been found as a symmetry
of the quantum Hall system \cite{trug}.
This symmetry is generated by the
magnetic translation operators in the $x-y$ plane,
i.e. translations in y up to x-dependent gauge
transformations
(in a
gauge where the electromagnetic potential is
$A_0 =0, A_x=-By, A_y =A_z=0$ ) :
\bea
\nonumber
     b&\equiv &-\partial _y + i \partial _x + ieBx \\
     b^{\dagger}&\equiv &\partial _y + i \partial _x -ieBx
\label{mtr}
\eea

\nk where
$B$ denotes the external magnetic
field in the direction perpendicular to the plane.
Integer powers of the
operators (\ref{mtr})
\be
              V_{n,m}=(b^{\dagger})^{n+1}(b)^{m+1} \qquad :
\qquad n,m \ge -1
\label{walg}
\ee

\nk generate quantum deformations of $W$-algebras \cite{trug}
that
include
area-preserving diffeomorphisms of the type
recently argued \cite{emn1} to be responsible for quantum
coherence in systems with space-time singularities.
As we mentioned earlier,
the difference of the Hall systems discussed in \cite{trug}
from the ordinary $W_{\infty}$-case is the requirement of
regularity
of the wavefunction at the position of the electron. This
implies a {\it truncation} of the algebra to the positive
modes (\ref{walg}).
On the other hand
the KP-hierarchy
\be
\Lambda _{KP}=\partial _z + \sum _{i=0}^{\infty} u_i
(\partial _z)^{-i-1}
\label{kp}
\ee

\nk that generates
the full $W_{\infty}$ algebras \cite{wu}, contains
modes corresponding to the
negative integer powers of $b$ and $b^{\dagger}$ in (\ref{walg}),
which lead to divergences of the
electron wavefunction at the origin.
Such divergences are associated
with
topological defects on the plane, as is the case of
annular or doped Hall conductors
\footnote{We remind the reader
that a ``missing'' electron changes the Hilbert space
of the problem in a topologically non-trivial way \cite{dm}.}.
The existence of the full $W_{\infty}$ algebra can be
verified directly in those cases using an effective
gauge field theory description
of
the lowest Landau level by a pure
Chern-Simons gauge theory on a topologically non-trivial
space, which is equivalent to the infinite topological-mass limit
of a Maxwell-Chern-Simons theory \cite{kogan}.
The generators of the full $W_{\infty}$ symmetry in that case
can be expressed
in terms of the magnetic translation operators (\ref{mtr}) as
\be
 W_{n,{\bar n}}= exp(\frac{2\pi}{k}nb^{\dagger})
exp(-\frac{2\pi}{k}{\bar n}b)
\label{torus}
\ee

\nk where $n \equiv -i \tau n + im $, ${\bar n}$ is the complex
conjugate,
$n,m$ are integers (not necessarily positive) and
$\tau$ denotes a (complex)
modular parameter of the Riemann surface
on which
the system is defined. The coefficient of the Chern-Simons term
is normalized to $k / 8\pi $ and $[b,b^{\dagger}]
= k/2\pi Im\tau  $. Compared with the smooth case of ref. \cite{trug},
the Chern-Simons theory can be considered as describing excitations
on the edge
of a {\it large}
disk
which the Hall system lives on. In that case,
since the relevant wave-functions are defined far away
from the origin, any regularity requirements can be relaxed,
leading to an enhancement of the
symmetry to the full $W_{1+\infty}$.
\pr
We showed in ref. \cite{emntop} that a higher $W_{1+\infty} \otimes
W_{1+\infty}$ symmetry group is recovered at a space-time
singularity, as a bosonic subgroup of $N=2$ super
$W_{1+\infty}$ symmetry \cite{yu}. The spontaneous breakdown
of $W_{1+\infty}\otimes W_{1+\infty}$ to $W_{1+\infty}$
away from the singularity is accompanied by the appearance
of discrete massless leg-poles. In view of the
intimate connection of the Hall systems with $N=2$
supersymmetry \cite{gyro,halsusy},
we expect a similar
feature of the excitation spectrum in annular Hall
conductors : discrete long-range excitations
should appear whenever the planar topology of the Hall
conductor is non-trivial. The following facts point in this direction:
In Hall electron
systems
the $N=2$ supersymmetry is
generated by the
{\it covariant derivative } operators
\cite{halsusy}
\bea
 \nonumber a& \equiv & \partial _x + i\partial _y
               -ieBy            \\
 a^{\dagger}& \equiv &   -\partial _x + i\partial _y +ieBy
\label{covar}
\eea

\nk The latter
{\it commute}
with $b$ and $b^{\dagger}$,
but not with the Landau
Hamiltonian $H\propto a a^{\dagger} + a^{\dagger} a$, since
$[ a, a^{\dagger} ]=k/2\pi $.
These operators
generate \cite{kogan}
a $W_{\infty}$
algebra in a way similar to (\ref{torus}), which is not a
symmetry of the Landau
Hamiltonian.
Restriction to the first Landau level, make the effective gauge
theory infinitely massive, and thus purely topological. The
Landau
Hamiltonian formally
vanishes in that limit, and the $W_{\infty}$ algebra
generated by the
$a$ and $a^{\dagger}$ operators,
which acts on this
Landau level, may be promoted to a
symmetry algebra
of this reduced system.
In the
black hole case this
corresponds to the limit close to the singularity, where
there is an enhanced $W_{\infty} \otimes W_{\infty}$ \cite{emntop}.
\pr
This is just one example of the rich interplay that we expect
between studies of Hall conductors and string black holes
in view of the isomorphism we have developed. The above
paragraph is one example of a theoretical
discovery in black hole physics that may have implications
for experiments on Hall conductors. Conversely, measurements
on Hall conductors can be regarded as a laboratory for doing
experimental black hole physics on a {\it table top }.
\pr
It would be interesting to study the renormalisation group flow
that relates different fractions in the hierarchy of Hall systems,
which has been associated with extended duality symmetries
\cite{lutk}. It would seem that there is only one value
$k=\frac{9}{4}$ which admits a space-time interpretation in the
two-dimensional black hole case. However, the $W_{\infty}$-symmetry
structure of the WZ is essentially unchanged for other values
of $k$ \cite{bakir}. Moreover, we expect that it is possible
to construct consistent models by tensoring a WZ model with other
field theories on the world-sheet so as to obtain a total
central charge of $26$ (or $15$ for supersymmetric theories),
as required for the space-time interpretation.
Such mixed WZ models have recently been considered in connection
with higher-dimensional universes \cite{lust}. One would consider
such a tensored WZ model as corresponding to some Hall system
with a filling fraction falling in a hierarchy of values
appearing in a flow of central charge from the other tensored
components, interpreted as matter fields.
\pr
Looking at the form (\ref{triapente}) of the magnetic field
in the effective gauge theory description of the WZ model,
we see that it vanishes close
to the singularity, as well as at
the horizon. Thus the system close to the singularity
resembles that of a zero-field Hall conductor, as discussed
in section 1. This is consistent with the spontaneous nature
of $T-$ and $P-$ violation \cite{hlad}, which, we have earlier
argued, explains the irreversibility of black hole decay
and the arrow of time when applied to the cosmological singularity.
It is natural to ask whether space-time is still a Hall liquid far
away from the singularity, where the Chern-Simons description acquires
higher-order corrections. The answer is yes, according to recent
studies of $c=1$ matrix models \cite{indians}, which represent
the spatially-asymptotic form of the black hole. The character
of an incompressible Hall fluid and the associated
$W_{1+\infty}$-symmetry
structure are preserved \cite{grossnewm,indians}.
Thus space-time foam, as described by a system of multiple
topological defects on the world-sheet, shares many common features
with Hall fluids.

\section{Renormalisation Group Flow
and the Arrow of Time}
\pr
We have argued recently \cite{emnrg} that
space-time foam can be formulated as a renormalization
flow problem on the world-sheet, through the identification
of the renormalization group mass scale, that was introduced
as a covariant world-sheet cut-off, with the target time in
Planck units. This cut-off must be introduced to separate
the light degrees of freedom measured in laboratory experiments
from the massive string degrees of freedom which are not
observed. The massless states are related to the massive states
by $W$-symmetries which maintain quantum coherence \footnote{It
should be noticed that the $W$-symmetries pertain to the
two-dimensional target space string theory, where the
massive modes are solitonic non-propagating modes having
definite energies and momenta \cite{grossnewm,emn1}. Two-dimensional
strings have been argued to
constitute
the s-wave sector of higher-dimensional
target spaces \cite{emnfour}.} \cite{emn1}, and
quantum mechanics is modified in the effective theory of the
light degrees of freedom, which behave like an open system
with monotonically increasing entropy \cite{emnrg}.
It is worth emphasizing the fact that the full string theory,
with its massive degrees of freedom, is a conformally
invariant theory in which there is no renormalization group flow.
This in some sense defines a
concept of ``eternity'' within which the light-mode subsystem
(our world) interacts thermodynamically by exchanging
energy with its environment (massive string states). This
leads
to an
irreversible
time flow for an observer who is part of this
subsystem. Formally the above ideas can be expressed
as follows \cite{emnrg}. Let
$ \{ G^i  \} $  be a collective notation for
the
heavy and light
string modes,
and $t$ be
the renormalisation group scale.
The
evolution equation of
the density matrix $\rho $
away from a (conformally invariant) fixed point reads \cite{emnrg}
\be
   \dot \rho \equiv
   \frac{\partial \rho}{\partial t } = i[ \rho , H ] + iG_{ij}
[\rho , G^i]\beta^j
\label{rgo}
\ee

\nk where $H$ is the Hamiltonian and $\beta^i =dG^i/dt$  is the
renormalization group $\beta $-function.
The non-triviality of the friction non-commutator
term in (\ref{rgo}) {\it implies } the non triviality
of the commutator term. Indeed, for systems interacting
with a reservoir of particles at
temperature $\beta ^{-1}$, as is our light-mode system,
the density
matrix is expressed as $exp(\beta(F-H))$
where F is the free energy. In the case of strings
$F$ is the generating functional of connected string amplitudes
and is associated \cite{mmeff}
with the Zamolodchikov c-function \cite{zam}.
At a conformally-invariant
 fixed point (fp) the latter becomes a c-number,
the central charge of the theory, and so $[\rho , H ]_{fp}=0$,
and the evolution equation (\ref{rgo}) has trivial
content. On the other hand, by considering deformations
of the pertinent stringy $\sigma$-models by massless states only
\cite{emnrg}, one goes away from the conformally-invariant fixed point
in an irreversible way, as implied by Zamolodchikov's c-theorem
\cite{zam,mavmir}. The situation is similar to the conventional
renormalization group flow in local field theories.
Due to logarithmic infinities, the experimentalist
in the laboratory measures ``running'' coupling constants
correponding to light particles. Formally, the light
mode infinities can be remedied by introducing heavy states
which make the theory finite and are responsible for the
renormalization group flow of the light states.
This formal procedure of
local field theories acquires, therefore,
an important
physical meaning in string theory propagating
in singular backgrounds,
such as black holes or cosmological singularities
\cite{witt,lust}.
Within the framework
of string theory the observed flow of time is a
consequence \cite{emnrg}
of the existence of massive (solitonic)
string modes which couple to the observed light states
on account of the $W$-symmetries \cite{emn1}.
The irreversibility of the renormalization group flow
implied by Zamolodchikov's c-theorem \cite{zam,mavmir}
corresponds to the $P-$ and $T-$ violation in the effective
Chern-Simons theory of the Hall fluid.
\pr
This $P-$ and $T-$ violation
also determine the cosmological arrow of time, as we
now argue. In the Hall analogue description of
singular space-times, one introduces a third
dimension $\tau$, which plays the r\^ ole of an
adiabatic evolution parameter. As explained
in \cite{emnd}, $\tau$ can be interpreted as an inverse
pseudo-``temperature'' which determines the phase
structure of the universe through a BKT transition
\cite{kt} from a hot Euclidean to a cold Minkowski
space-time. The irreversibility of this cosmological
evolution in pseudo-``temperature'' is guaranteed
by the $T-$violation induced by the three-dimensional
Chern-Simons theory. The associated $P-$violation implies
reflection non-invariance on the world-sheet which the
embedding (\ref{triaexi}) elevates into target-time
irreversibility. The latter is associated
with the irreversibility of the renormalization group flow
once the world-sheet cut-off mass-scale is identified with
the target time.
\pr
Thus the
renormalization group flow down the Hall hierarchy \cite{lutk}
and the $P-$ and $T-$ violation in the Hall vacuum correlate
the cosmological arrow of time, the decay of black holes,
and a monotonic increase in entropy at the microscopic and
macroscopic levels.
We plan to return in a future paper \cite{emnap} to a
more quantitative study of this and other cosmological issues
in string theory.
\newpage
\pr
\nk {\Large  {\bf Acknowledgements}}
\pr
We acknowledge a useful discussion with A. Cappelli and G. Zemba.
The work of D.V.N. is partially supported by DOE grant
DE-FG05-91-ER-40633 and by a grant from Conoco Inc.

\end{document}